\documentclass[runningheads,a4paper]{llncs}

\usepackage{amssymb}
\usepackage{graphicx}

\usepackage{url}
\urldef{\mailsa}\path|{x.d.arsiwalla@gmail.com}| 
\newcommand{\keywords}[1]{\par\addvspace\baselineskip
\noindent\keywordname\enspace\ignorespaces#1}

\begin{document}

\mainmatter  

\title{The Global Dynamical Complexity of the Human Brain Network } 


%
%
\author{Xerxes D. Arsiwalla\inst{1}%
\and Paul Verschure\inst{1,2}}
%

\institute{Synthetic Perceptive Emotive and Cognitive Systems (SPECS) Lab, \\ Center of Autonomous Systems and Neurorobotics, \\ Universitat Pompeu Fabra, \\  
Barcelona, Spain.\\ 
\and
Instituci\'{o} Catalana de Recerca i Estudis Avan\c{c}ats (ICREA), \\ 
Barcelona, Spain.\\
\mailsa\\ 
}

%
%


\pagestyle{empty}
\maketitle

\begin{abstract}
How much information do large brain networks integrate as a whole over the sum of their parts? Can the dynamical complexity of such networks be globally quantified in an information-theoretic way and be meaningfully coupled to brain function? Recently, measures of dynamical complexity such as integrated information have been proposed. However, problems related to the normalization and Bell number of partitions associated to these measures make these approaches computationally infeasible for large-scale brain networks. Our goal in this work is to address this problem. Our formulation of network integrated information is based on the Kullback-Leibler divergence between the multivariate distribution on the set of network states versus the corresponding factorized distribution over its parts. We find that implementing the maximum information partition optimizes  computations. These methods are well-suited for large networks with linear stochastic dynamics. We compute the integrated information for both, the system's attractor states, as well as non-stationary dynamical states of the network.  We then apply this formalism to brain networks to   compute the integrated information for the human brain's connectome. Compared to a randomly re-wired network, we find that the specific topology of the brain generates greater information complexity. 

\end{abstract}

\section{Introduction}

From a computational neuroscience perspective, the brain is oftentimes abstracted as a complex information processing network, that integrates sensory inputs from multiple modalities in order to generate action and cognition. In this paper, we ask a much simpler question: viewing the brain as a dynamical network of neural masses, how can one compute the information integrated by such networks in the course of dynamical transitions from one state to another? A possible approach, among others, is to look at information-theoretic complexity measures that seek to quantify information generated by all causal sub-processes in such a network. One candidate measure for global dynamical complexity is integrated information, usually denoted as $\Phi$. It was first introduced in neuroscience as a complexity measure for neural networks, and by extension, as a possible correlate of consciousness itself  \cite{tononi1994}. It is defined as the quantity of information generated by a network as a whole,  due to its causal dynamical interactions, and one that is over and above the information generated independently by the disjoint sum of its parts.  As a complexity measure, $\Phi$  seeks to operationalize the intuition   that complexity arises from simultaneous integration and differentiation of the network's structural and dynamical properties. As such, the interplay of  integration and differentiation in a network's dynamics is hypothesized to generate information that is highly diversified yet integrated, thereby creating patterns of high complexity.  The aim of this paper is to develop mathematical tools for computing integrated information (analytically when possible, otherwise numerically) for large networks.  We then  apply this framework to the large-scale structural connectivity network of the human brain. 

Let us begin with a brief review of the rich history of this field. The earliest proposals defining integrated information were made in the pioneering work of \cite{tononi1994}, \cite{tononi2003} and \cite{tononi2004}.  Since then, considerable progress has been made towards development of a normative theory as well as applications of integrated information  \cite{bt}, \cite{balduzzi2009qualia}, \cite{bs}, \cite{tononi2012integrated},  \cite{arsiwalla2013iit}, \cite{oizumi2014},  \cite{arsiwalla2016computing}, \cite{mediano2016integrated}, \cite{krohn2016computing}.  Similar information-based approaches have also been successfully applied to many-body problems in other domains, such as, for the problem of estimating microstates of statistical mechanical ensembles \cite{arsiwalla2009entropy}. In fact, there are now several candidate measures of integrated information such as neural complexity \cite{tononi1994}, causal density \cite{seth2005causal}, $\Phi$ from integrated information theory: IIT 1.0, 2.0 \& 3.0 \cite{tononi2004,bt,oizumi2014}, stochastic interaction \cite{wennekers2005stochastic,ay2015information}, empirical $\Phi$ \cite{bs} and synergistic $\Phi$ \cite{griffith2014quantifying,griffith2014principled}, plus several variations of these (see \cite{tegmark2016improved} for an overview). Table \ref{T1} summarizes these measures along with corresponding information metrics upon which they have been based.  

\begin{table}[h!] 
\centering 
\caption{Candidate measures of integrated information shown alongside information metrics used in their respective formulations.  }
\label{T1}
      \begin{tabular}{|c|c|}
        \hline
        $\Phi$ Measures  & Information Metrics  \\ \hline      
        Neural Complexity & Mutual Information \\
        Causal Density & Granger Causality \\ 
        Stochastic Interaction & Kullback-Leibler Divergence  \\
        IIT 1.0 \& 2.0 & Kullback-Leibler Divergence  \\
        Empirical $\Phi$ & Mutual Information \\   
        IIT 3.0 & Earth Mover's Distance \\       
        Synergistic $\Phi$ & Synergistic Information  \\ \hline
      \end{tabular}
\end{table}

Many of the above measures have been useful in different domains of validity. However, applications to realistic data and in particular to large-scale networks have proven  computationally challenging. With a focus on developing computational tools, we discuss three of the above measures in more detail. The measure of \cite{bt} has been quite useful for discrete-state, deterministic, Markovian systems with the maximum entropy distribution. On the other hand, the measure of \cite{bs}  has been applied to continuous-state, stochastic, non-Markovian systems and in principle, admits dynamics with any empirical distribution (although in practice, it is easier to use assuming Gaussian distributions). The formulation in \cite{bs} is based on mutual information, whereas \cite{bt} uses a measure based on the Kullback-Leibler  divergence. Note however, that in some cases the measure of \cite{bs} can take negative values and that complicates its interpretation. The Kullback-Leibler based definition  computes the information generated during state transitions and as we shall see remains positive in the regime of stable dynamics. This makes it easier to interpret as an integrated information measure. Both measures \cite{bt}, \cite{bs} make use of a normalization scheme in their formulations. Normalization inadvertently  introduces ambiguities in computations. The normalization is actually used for the purpose of determining the partition of the network that minimizes the integrated information, but a normalization dependent choice of partition ends up influencing the value and interpretation of  $\Phi$.  An alternate measure based on the Earth  Mover's distance was proposed in \cite{oizumi2014}. This does away with the normalization  problem (though the current version is not formulated for continuous-state variables). However,  the formulation of  \cite{oizumi2014}  lies outside the scope of standard information theory and  is still very difficult for performing computations on large networks. 

A  comment on network partitions is relevant at this point. The three measures of $\Phi$ discussed above, make use of what is called the minimum information partition or minimum information bi-partition (MIP/MIB). The issue with this partitioning is that it leads to a  combinatorial explosion in the number of configurations to be evaluated when working with networks having a large number of nodes. As a result, application of the above measures to  compute integrated information of very large networks remains challenging, particularly for the scale of networks obtained from neuroimaging data. On the other hand, in earlier work  \cite{arsiwalla2013iit}, we have introduced a formulation of integrated information that overcomes both, the normalization and combinatorial problem by using a different partitioning of the network called the maximum information partition (MaxIP), which opens the prospect of large-scale applications. However, the formulation in  \cite{arsiwalla2013iit}  was only applicable for uncorrelated node dynamics, which may not be  realistic enough for many biological systems. 

In this paper, we seek to go beyond \cite{arsiwalla2013iit,arsiwalla2016computing,arsiwalla2016high}, starting with an extension of the formalism to include node correlations and also non-stationarity. In order to do that, we solve the discrete-time Lyapunov equation, the solution of which, is then used to get fully analytic expressions for $\Phi$ with network correlations. We consider networks with linear stochastic dynamics, which generate multivariate time-series signals. Furthermore, our networks are plastic, in the sense that connection weights are scalable using a global coupling parameter. We compute $\Phi$ as a function of this coupling. We also extend our framework to include non-stationary dynamics. This gives us $\Phi$ as a function of time, computed through the temporal evolution of the system. The stationary solution yields $\Phi$ at the fixed-point attractor, whereas the non-stationary solution leads to $\Phi$ elsewhere in the phase space of the system. 

As proof of principle, we apply our formulation to the structural connectivity network of white matter fiber tracts in the human cerebral cortex, obtained from diffusion spectrum imaging \cite{Hagmann2008}, \cite{honey2009predicting}. This network has 998 nodes, representing neuronal populations. The edges are weighted fiber counts between populations. Implementing stochastic Gaussian dynamics on this network, we determine stationary solutions to the dynamical system from which we compute the information integrated in bits. To contrast with a null-model, we randomly re-wire the original network and repeat the computation. The original network scores higher on integrated information for all allowed couplings in the stationary as well as non-stationary regime.

\section{Stochastic Integrated Information}
\subsection{Mathematical Formulation} 
We consider networks with linear stochastic dynamics. The state of each node is given by a random variable pertaining to a given probability distribution. These variables may either be discrete-valued or continuous. However, for many biological applications,  Gaussian distributed, continuous-valued state variables are fairly reasonable abstractions (for example, aggregate neural population firing rate, EEG or  fMRI signals). The state of the network ${\bf X_t}$ at time $t$ is taken as a multivariate Gaussian variable with distribution ${\bf P_{X_t} (x_t) }$. ${\bf  x_t}$ denotes an instantiation of ${\bf X_t}$ with components ${ x_t^i }$ ($i$ going from $1$ to $n$, n being the number of nodes). When the network makes a transition from an initial state  ${\bf X_0}$ to a state ${\bf X_1}$ at time $t = 1$, observing the final state generates information about the system's initial state.  The information generated equals the reduction in uncertainty regarding the initial state ${\bf X_0}$. This is given by the conditional entropy ${ \bf H (X_0 | X_1) }$. In order to extract that part of the information generated by the system as a whole, over and above that generated individually by its parts, one computes the relative conditional entropy  given by the Kullback-Leibler divergence of the conditional distribution ${\bf P_{X_0 | X_1 = x'} (x) }$ of the system with respect to the joint conditional distributions $\prod_{k=1}^{r}  {\bf P_{M^k_0 | M^k_1 = m'}  }$ of its non-overlapping sub-systems demarcated with respect to a partition ${\cal P}_r$ of the system into $r$ distinct sub-systems. Denoting this as ${ \Phi_{ {\cal P}_r  } }$, we have    
\begin{eqnarray}
{ \Phi_{ {\cal P}_r  } }  ( {\bf X_0 \rightarrow X_1 = x'}  )   =  \,  D_{KL} \left(   {\bf P_{X_0 | X_1 = x'}  }  \big|\big|  \prod_{k=1}^{r}  {\bf P_{M^k_0 | M^k_1 = m'}  }  \right) 
\label{KLdiv}
\end{eqnarray}
where for an $r$ partitioned system, the state variable ${\bf X_0}$ can be decomposed as a direct sum of state variables of the sub-systems 
\begin{eqnarray}
{\bf  X_0 = M_0^1 \oplus M_0^2 \oplus  \cdots \oplus M_0^r =  \bigoplus_{k = 1}^{r}  M_0^k  }
\end{eqnarray}
and similarly,  ${\bf X_1}$ decomposes as 
\begin{eqnarray}
{\bf  X_1 = M_1^1 \oplus M_1^2 \oplus  \cdots \oplus M_1^r =  \bigoplus_{k = 1}^{r}  M_1^k  }
\end{eqnarray}  
For stochastic systems, it is useful to work with a measure that is independent of any specific instantiation of the final state ${\bf x'}$. So we average with respect to final states to obtain an expectation value from eq.(\ref{KLdiv}). After some algebra, we get   
\begin{equation}
\left<   \Phi  \right>_{{\cal P}_r  }  ( {\bf X_0 \rightarrow X_1 }  )  =   -   {  \bf H  (X_0 | X_1) }  +   \sum_{k=1}^{r}  { \bf H (M^k_0 | M^k_1) }
\label{infointr}
\end{equation}
This is our definition of integrated information, which we use in the rest of this paper. Note that the measure described in \cite{bt} is not applicable to networks with stochastic dynamics.  They do use eq.(\ref{KLdiv}) as their definition but endow their nodes with discrete states. On the other hand,  \cite{bs} uses a different definition of integrated information, where conditional entropies as in eq.(\ref{infointr}) are replaced by conditional mutual information. This definition only matches the definition of  eq.(\ref{KLdiv})  in special cases but not in general for any distribution. From an information theory perspective, the Kullback-Leibler divergence offers a principled way of comparing probability distributions, hence we follow that approach in formulating our measure in eq.(\ref{infointr}).  

The state variable at each time $t = 0$ and  $t = 1$ follows a multivariate Gaussian distribution   
\begin{equation}
{\bf X_0  \sim   {\cal N} \left(  \bar{x}_0, \Sigma (X_0)  \right)   }    \qquad   {\bf X_1  \sim   {\cal N} \left(  \bar{x}_1, \Sigma (X_1)  \right)   }    
\end{equation}
The generative model for this system is equivalent to a multi-variate auto-regressive process      \cite{barrett2010}        
\begin{equation}
{\bf X_1 =  {\cal A} \;  X_0  + E_1 }   
\label{genmod}
\end{equation}
where ${\cal A}$ is the weighted adjacency matrix of the network and $E_1$ is Gaussian noise.  Next, taking the mean and  covariance respectively on both sides of this equation, while holding the residual independent of the regression variables, yields 
\begin{eqnarray}
{\bf   \bar{x}_1  =    {\cal A} \;    \bar{x}_0   }  \quad  \qquad   {\bf  \Sigma(X_1)  =   {\cal A}  \;   \Sigma(X_0)  \;  {\cal A}^T  +  \Sigma(E)    }     
\label{DTLeq}   
\end{eqnarray}
In the absence of any external inputs, stationary solutions of a stochastic linear dynamical system as in eq.(\ref{genmod}) are fluctuations about the origin. Therefore, we can shift coordinates to set the means ${\bf  \bar{x}_0 }$ and consequently ${\bf  \bar{x}_1 }$ to the zero. The second equality in eq.(\ref{DTLeq}) is the discrete-time Lyapunov equation and its solution will give us the covariance matrix of the state variables. 

The conditional entropy for a multivariate Gaussian variable was computed in \cite{bs} 
\begin{equation}
{\bf H (X_0 | X_1)  }  =   \frac{1}{2}  n  \log (2 \pi e)  - \frac{1}{2}   \log \left[  \det    {\bf \Sigma (X_0 | X_1) }  \right]   
\end{equation}
which is fully specified by the conditional  covariance matrix. Inserting this in eq.(\ref{infointr})  yields 
\begin{equation}
\left<   \Phi  \right>_{{\cal P}_r }  ( {\bf X_0 \rightarrow X_1 }  )   =  \frac{1}{2}   \log \left[  \frac{ \prod_{ {\bf k} = 1}^{r}  \det   {\bf \Sigma (M^k_0 | M^k_1) }  }{ \det    {\bf \Sigma (X_0 | X_1) }  }  \right]    
\label{ginfo}
\end{equation}
Now, in order to compute the conditional covariance matrix we make use of the identity  (proof of this identity for the Gaussian case was demonstrated in \cite{barrett2010}) 
\begin{equation}
 {\bf \Sigma (X | Y)  =  \Sigma(X) -  \Sigma (X,  Y)  \Sigma (Y)^{-1}  \Sigma (X, Y)^T    }    
\label{covid}
\end{equation}
The appropriate covariance we will need to insert in this expression is  
\begin{equation}
{\bf \Sigma (X_0,  X_1)  \equiv \left<  \left( X_0 - \bar{x}_0 \right)  \left( X_1 - \bar{x}_1  \right)^T  \right>  =  \Sigma (X_0)  \,  {\cal A}^T   }     
\label{X01cov}
\end{equation}
which gives for the conditional covariance 
\begin{eqnarray}
{\bf \Sigma (X_0 | X_1)  =  \Sigma(X_0)  -  \Sigma (X_0)  \,  {\cal A}^T \,  \Sigma (X_1)^{-1}    {\cal A}  \;  \Sigma (X_0)^T  }    
\label{Xcondcov}
\end{eqnarray}
And similarly for the sub-systems
\begin{eqnarray}
 {\bf \Sigma (M^k_0 | M^k_1) }   =  {\bf   \Sigma(M_0^k) }  -   {\bf  \Sigma(M_0^{k})  \,  {{\cal A}^T} \big{|}_{k}  \,  { \Sigma(M_1^{k})  }^{-1}   {\cal A}  \big{|}_{k}  \,  {\Sigma (M_0^k) }^T    }
\label{Mcondcov}
\end{eqnarray}
where $k$ indexes the partition such that ${\bf M_0^k}$ denotes the $k^{th}$ sub-system at $t = 0$ and $ {\cal A}  \big{|}_{k}$ denotes the restriction of the adjacency matrix to the $k^{th}$ sub-network.  

Further, for linear multi-variate systems, a unique fixed point always exists. We try to find stable stationary solutions of the dynamical system. In that regime, the multi-variate probability distribution of states approaches stationarity and the  covariance matrix converges, such that 
\begin{eqnarray}
{\bf \Sigma (X_1) = \Sigma (X_0)}
\label{covstat}
\end{eqnarray}
$t = 0$ and $t = 1$ refer to time-points taken after the system converges to the fixed point. Then the discrete-time Lyapunov equations can be solved iteratively for the stable covariance matrix ${\bf \Sigma (X_t)}$. For networks with symmetric adjacency matrix and independent Gaussian noise, the solution takes a particularly simple form
\begin{eqnarray}
{\bf \Sigma (X_t)  =  \left( 1 -  {\cal A}^2 \right)^{-1}  \Sigma(E)  }      
\label{covsol}
\end{eqnarray}
and for the parts, we have
\begin{eqnarray}
{\bf  \Sigma(M_0^k)  =  \Sigma (X_0)  \big{|}_{k}    }      
\label{covparts}
\end{eqnarray}
given by the restriction of the full covariance matrix on the $k^{th}$ sub-network. Note that eq.(\ref{covparts}) is not the same as eq.(\ref{covsol}) on the restricted adjacency matrix as that would mean that the sub-network has been explicitly severed from the rest of the system. Indeed,  eq.(\ref{covparts}) is precisely the covariance of the sub-network while it is still part of the network  and $\left< \Phi \right>$ yields the integrated and differentiated information of the whole network that is greater than the sum of these connected parts.  Inserting eqs.(\ref{Xcondcov}), (\ref{Mcondcov}), (\ref{covsol}) and (\ref{covparts}) into eq.(\ref{ginfo}) yields $\left< \Phi \right>$ as a function of network weights for symmetric and correlated networks. For the case of asymmetric weights, the entries of the covariance matrix cannot be explicitly expressed as a matrix equation. However, they may still be solved by Jordan decomposition of both sides of the Lyapunov equation.

\subsection{The Maximum Information Partition}
Following \cite{edlund2011integrated} and \cite{arsiwalla2013iit},   the maximum information partition (MaxIP) is defined as the partition of the system into its irreducible parts. This is the finest partition and is  unique as there is only one way to combinatorially reduce a system into all of its sub-units. This partition can directly be found by construction and does not require a   normalization scheme for  sampling  through the space of multi-partitions in order to search for the one that either maximizes or minimizes the integrated information.  Consequently, the resulting  value of $\left< \Phi \right>$ computed using the MaxIP is free from normalization dependencies.        

Moreover, the MaxIP also helps reduce computational cost. This can be seen as follows. Prescriptions using the MIP/MIB are typically evaluated for a large class of network bi-partitions, whereas the MaxIP is uniquely defined. The number of bi-partitions of a set of $n$ elements is given by the sum of binomial coefficients $\sum_{p = 1}^{[n/2]}   \, ^nC_p$, where $^nC_p = n! / p ! \; ( n - p ) !$ with $n ! = n \times (n-1) \times \cdots \times 1$  and  $[n/2]$ denotes the nearest integer less that or equal to $n/2$. Among all possible bi-partitions, MIP/MIB prescriptions usually restrict to those that divide the system into approximately equal parts. This still leaves us with  $^n C_{[n/2]}$ configurations for which $\left< \Phi \right>$ has to be computed. Table \ref{T2} summarizes how this number scales with network size from a single node to a million nodes. 

\begin{table}[h!]
\centering 
\caption{Scaling of network configurations  upon computing $\Phi$ using the MIP/MIB versus using the MaxIP  for networks with $n$ nodes. } 
\label{T2} 
      \begin{tabular}{|c|c|c|}
        \hline
No. of nodes $n$  &  No. of equal part bi-partitions  $= \,  ^n  C_{[n/2]}$  & No. of  MaxIPs  $=  \, ^n   C_n$   \\  
\hline
1 & 1 & 1  \\
10   & 252  & 1  \\
100  & 1.01 $\times$ $10^{29}$ &  1  \\
1000  & 2.70 $\times$ $10^{299}$   &   1  \\
1000000  & 7.90 $\times$ $10^{301026}$    &   1  \\
\hline 
      \end{tabular}
\end{table}

Another interesting feature of the MaxIP is that $\left< \Phi \right>$ computed using this partition in fact accounts for the maximum amount of information that the network can integrate compared to any other bi-, tri- or multi-partition of the system.  This is due to the fact that this partition cannot be decomposed further. Every other partition will be coarser than the MaxIP and will therefore have at least some of its parts as composites of the irreducible units in the MaxIP. As these composites integrate more information than its own irreducible units, subtracting the information of a  composite (when treating the composite as a part) from the information of the whole system will always produce a smaller $\left< \Phi \right>$ than that obtained by subtracting the information of each irreducible unit of the network from that of the whole network. Therefore $\left< \Phi \right>$ computed using the MaxIP is the maximum possible integrated information of the system compared to $\left< \Phi \right>$ computed using any other partition of the network.  In that sense, unlike the MIP or MIB, the MaxIP in fact captures the complete information integrated by the network and is therefore a more natural choice for quantifying whole versus parts.

\section{Analytic Solutions for $\left< \Phi \right>$ }
\begin{figure}[h]
\centering
\includegraphics[height=5.0cm]{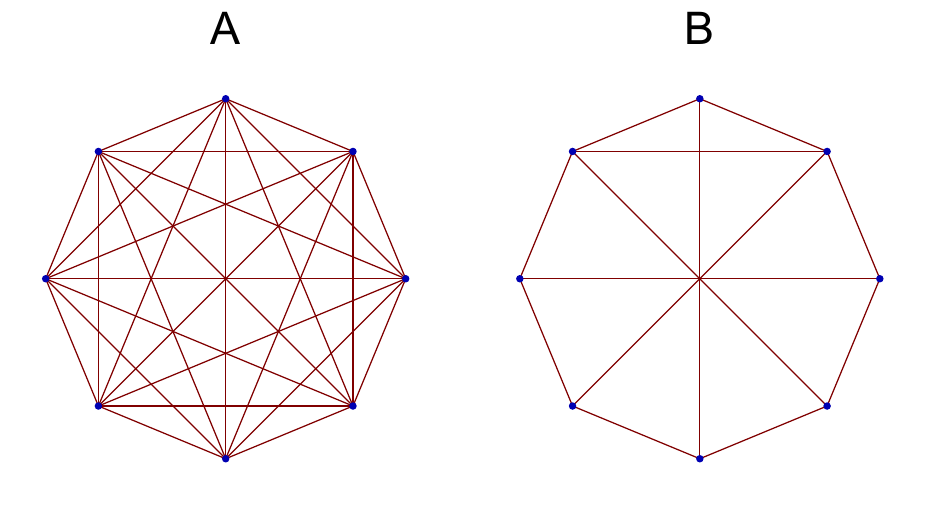}
\caption{{\bf Graphs of two artificial networks, (A) and (B).} }
\label{fig1a}
\end{figure}

Now that we have a rigorous analytic formulation of integrated information, let us first demonstrate examples of computations performed using artificial networks. In Figure~\ref{fig1a} we consider two artificial networks. For these cases, we want to compute the exact analytic solution for $\left< \Phi \right>$. Each of these networks have 8 dimensional adjacency matrices with bi-directional weights (though our analysis does not depend on that and works as well with directed graphs). We want to compute $\left< \Phi \right>$ as a function of network weights, which we keep as free parameters. However, in order to constrain the space of parameters, we shall set all weights to a single parameter, the global coupling strength $g$. This gives  us $\left< \Phi \right>$ as an analytic function of $g$.  

\subsection{$\left< \Phi \right>$ for Attractor States}
We first compute $\left< \Phi \right>$ in the stationary regime, that is, when the system has converged to its fixed-point attractor state. The results for the two networks labeled  A and B respectively are shown in eqs.(\ref{pnetA}), (\ref{pnetB})   respectively.  These are computed for a single time-step, corresponding to the stable stationary solution of the system.        

\begin{eqnarray}
\left<  \Phi  \right>_A &=& \frac{1}{2} \log \frac{ \left(1-43 g^2 \right)^8 }{ \left(1-50 g^2+49 g^4 \right)^8 }  \label{pnetA}  \\ 
\left<   \Phi  \right>_B &=& \frac{1}{2} \log \frac{B_1 \cdot B_2 \cdot B_3 \cdot B_4 \cdot B_5}{\left(-1+g^2 \right)^{4}  \left(1-8 g^2+4 g^4 \right)^{6} \left(1-17 g^2+72 g^4-64 g^6+16 g^8 \right)^{8} } \qquad 
\label{pnetB}   
\end{eqnarray}
where
\begin{eqnarray}
B_1 &=&  \left(1-15 g^2+56 g^4-56 g^6+16 g^8 \right)  \nonumber  \\
B_2 &=&  \left(1-15 g^2+54 g^4-54 g^6+16 g^8 \right)   \nonumber  \\
B_3 &=&  \left(1-22 g^2+159 g^4-426 g^6+336 g^8-80 g^{10} \right)^2   \nonumber  \\
B_4 &=&  \left(1-21 g^2+147 g^4-401 g^6+374 g^8-136 g^{10}+16 g^{12} \right)^2   \nonumber  \\ 
B_5 &=&  \left(1-23 g^2+183 g^4-612 g^6+835 g^8-526 g^{10}+152 g^{12}-16 g^{14} \right)^2   \nonumber  
\end{eqnarray}

Note that the mathematical framework described above is in no way limited by the size of the network and thus,  in principle,  can be applied to networks of any size, to yield exact results. The only practical difficulty would be in the form of available computing hardware resources. Hence, for very large data networks, such as those from brain imaging, numerical computations of $\left< \Phi \right>$ would be more practical to perform.

\subsection{$\left< \Phi \right>$ for Non-Stationary Dynamics }
The mathematical formulation developed above can also be used to compute $\left< \Phi \right>$ at non-stationary points in the solution space of networks with linear stochastic dynamics. We show this explicitly for Network B in Fig.~\ref{fig1a}. We compute $\left< \Phi \right>$ for the complete temporal evolution of the system starting from an initial condition at $t = 0$ until the system stabilizes at the fixed point attractor.  In the non-stationary case, eqs.(\ref{covstat}), (\ref{covsol}) and (\ref{covparts}) no longer hold. However, everything up to and including eq.(\ref{Mcondcov}) are valid. Hence, the covariance matrix ${\bf \Sigma (X_t)}$ is simply computed recursively following eq.(\ref{DTLeq}).  Subsequently,  ${\bf \Sigma (X_t)}$ and $\left< \Phi \right>$ are both computed for each time-point $t$. 

\begin{figure}[h!]
\begin{center}
\begin{minipage}[h]{0.5\linewidth} 
\centering
\includegraphics[width=0.97\linewidth]{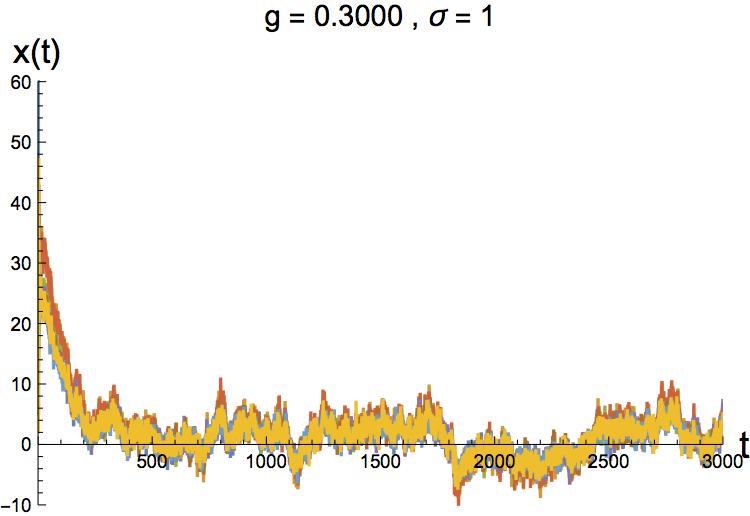} 
\end{minipage}%
\begin{minipage}[h]{0.5\linewidth} 
\centering
\includegraphics[width=0.97\linewidth]{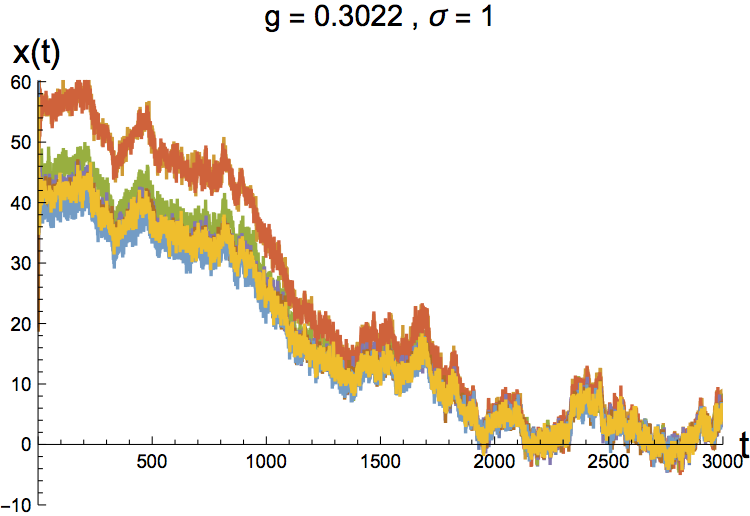} 
\end{minipage}
\end{center}
\caption{{\bf Simulated time-series data for Network B following the generative model in eq.(\ref{genmod}).}
      The plot on the left shows simulated data corresponding to network coupling strength $g = 0.3000$ and variance of noise $\sigma = 1$. The plot on the right refers to data from the same network with coupling $g = 0.3022$ and the same noise amplitude. Each plot shows 8 time-series profiles,  corresponding to the 8 nodes of the network (note that several of these profiles intersect or overlap with each other, hence in the above plots they appear to be clustered together). The time-series for each node is shown in a different color and the color scheme is the same for the plot on the left and the one on the right.  Stability of the system is guaranteed until the critical coupling at $g = 0.3023$. Closer to the critical point, the system takes longer to converge to the fixed-point attractor at ${\bf x = 0}$.  }
\label{fig3}
\end{figure}

\begin{figure}[h!]
\centering
\includegraphics[height=5.6cm]{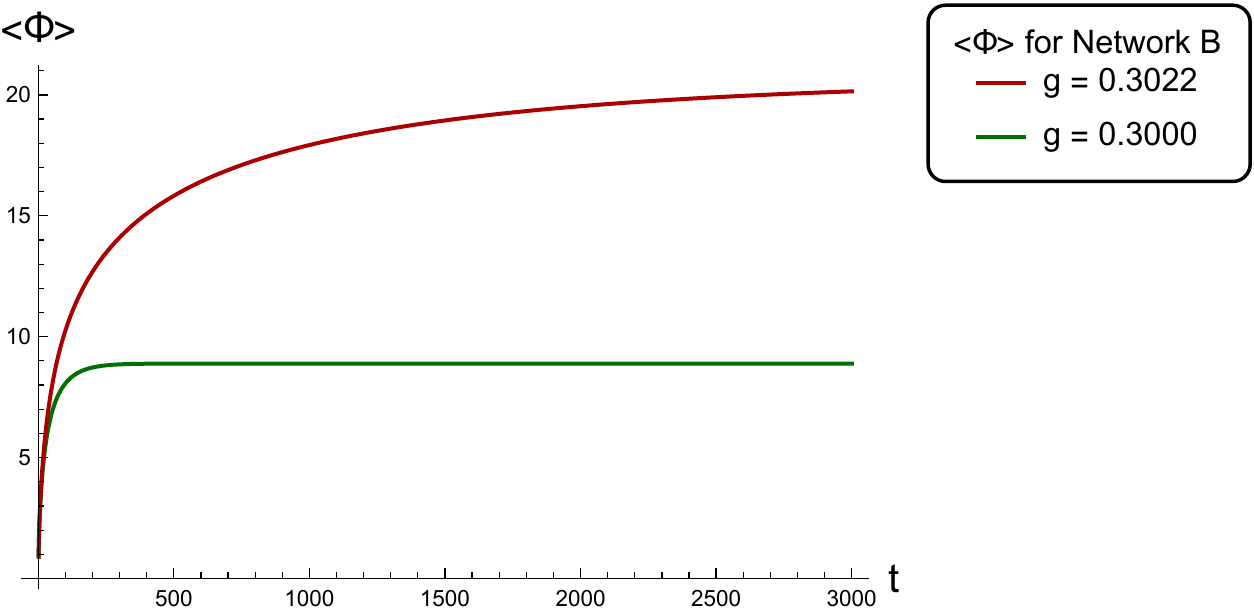}
\caption{{\bf Temporal profiles of $\left< \Phi \right>$ for Network B corresponding to the two coupling strengths used in Fig.~\ref{fig3}.}
        $\left< \Phi \right>$ saturates as the system approaches the stable attractor with greater integrated information for dynamics closer to the critical point.  }
\label{fig4}
\end{figure}

Fig.~\ref{fig3} shows the multivariate time-series signals generated by Network B for two different coupling strengths $g$. The critical value of $g$ for this network is $0.3023$, at which the dynamics becomes unstable. For $g \leq  0.3023$, the system converges to the fixed-point at the origin. In Fig.~\ref{fig4}  we plot temporal profiles of $\left< \Phi \right>$ for both the above values of $g$, which shows increasing integrated information for stronger coupled networks. 

\section{Application to Brain Connectomics}

\begin{figure}[h]
\begin{center}
\begin{minipage}[h]{0.45\linewidth} 
\centering
\includegraphics[width=0.99\linewidth]{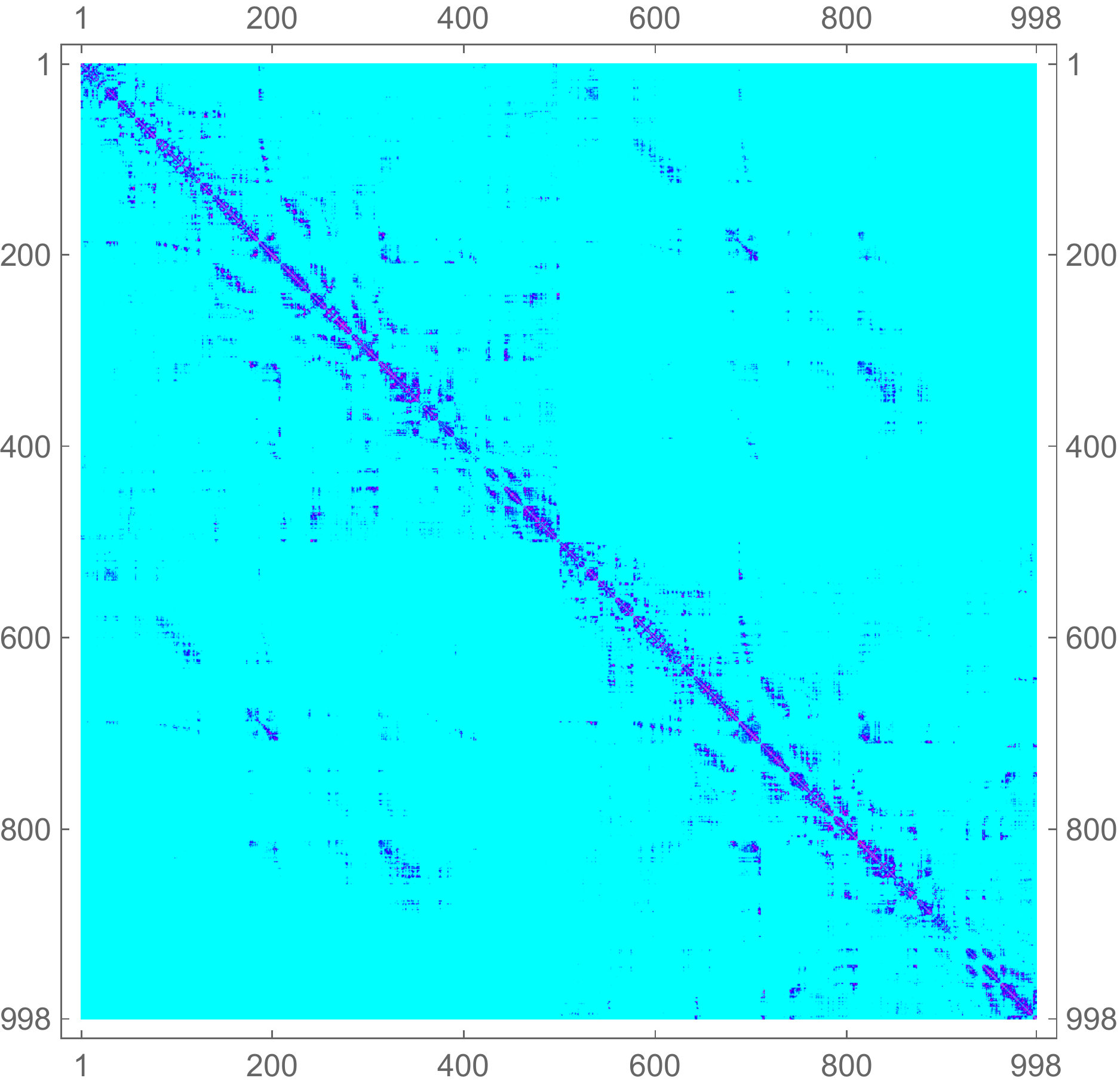} 
\end{minipage}%
\begin{minipage}[h]{0.1\linewidth} 
\centering
\includegraphics[width=0.7\linewidth]{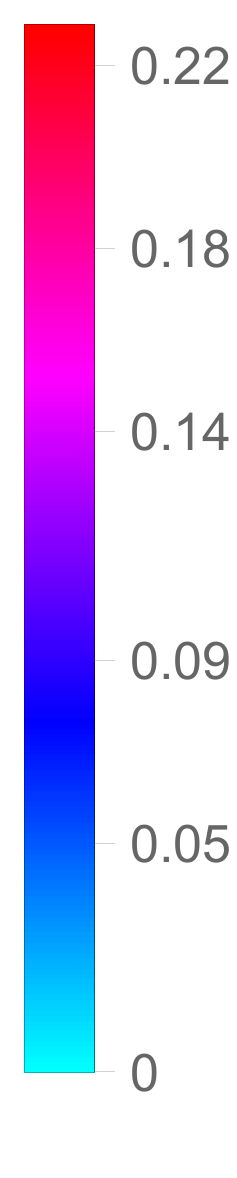} 
\end{minipage}%
\begin{minipage}[h]{0.45\linewidth} 
\centering
\includegraphics[width=0.99\linewidth]{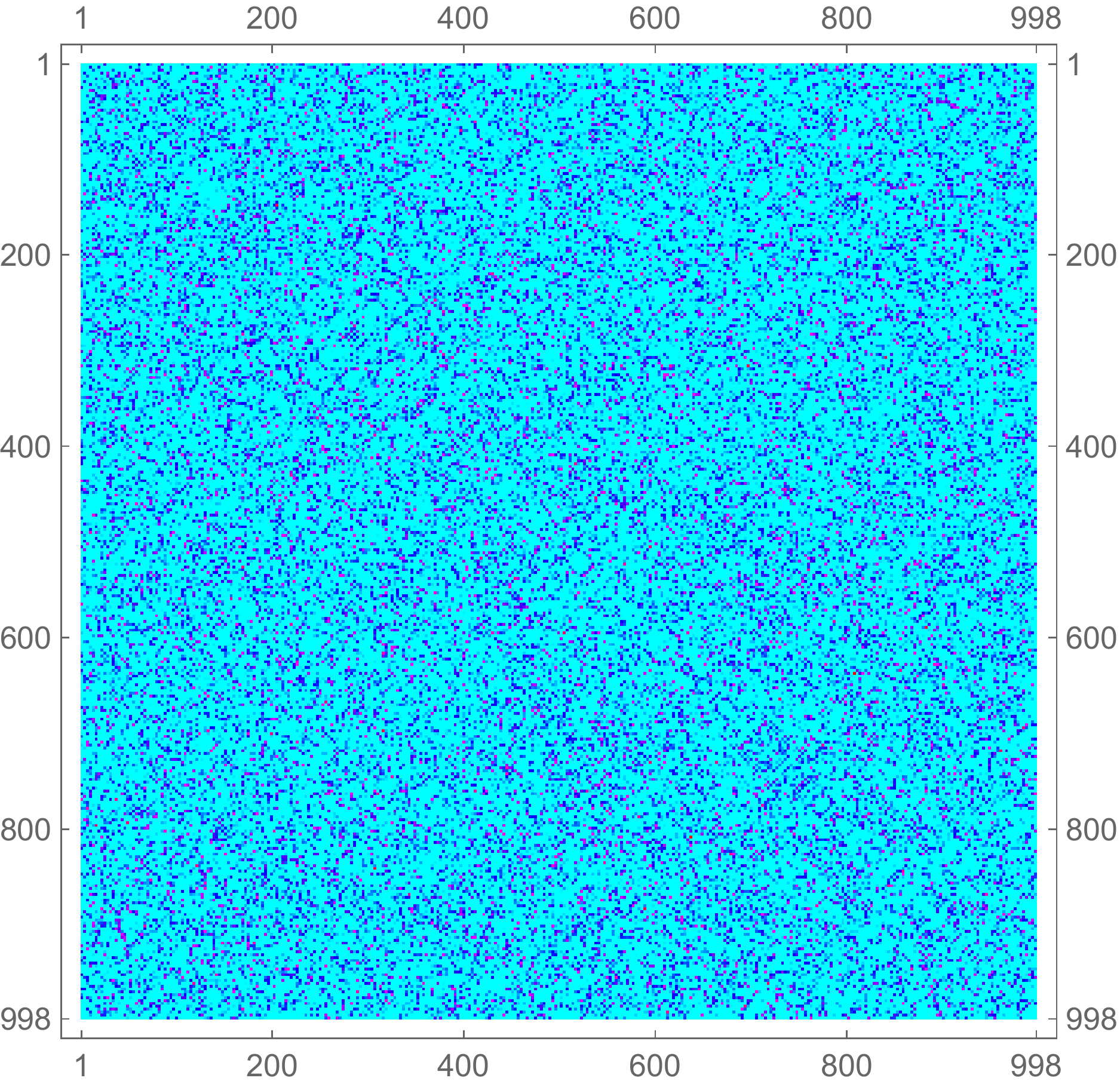} 
\end{minipage}
\end{center}
\caption{{\bf {\it Left}: Connectivity matrix of human cerebral white matter. {\it Right}: Randomized version of the same matrix, preserving network weights. }
  The data consists of white matter fiber tracts from 998 cortical voxels. The connectivity matrix on the left is a weighted matrix with the color-bar (in the middle) indicating connection strengths. The randomized matrix on the right is obtained by randomly shuffling positions of weights from the connectivity matrix. }
\label{fig1}
\end{figure}

The framework described above, provides us with all the mathematical tools to compute how much information is integrated in bits in a single time-step, by a large network with linear stochastic (Gaussian)  dynamics. We apply the above formulation to the whole brain structural connectivity network of the human cerebral cortex, using data published in the seminal work of \cite{Hagmann2008} and  \cite{honey2009predicting}. This data is acquired from high-resolution T1-weighted diffusion spectrum imaging (DSI). The data preprocessing pipeline, as described in \cite{Hagmann2008},  involves white and gray matter segmentation from the T1 images, followed by parcellation into 66 anatomical regions and subsequently 998 individual regions of interest (ROIs) based on Talairach coordinates. After that, whole brain tractography is performed to obtain estimates of axonal trajectories across the entire white matter. From this, connection weights between  pairs of ROIs are determined, resulting in a weighted network of structural connectivity across the entire brain. We have displayed the data in matrix form as a     998 dimensional matrix on the left-hand side of  Fig.~\ref{fig1}. The 998 voxels (ROIs) represent  nodes of the network. Each node is physically a population of neurons. The edges are weighted fiber counts between populations. Additionally, we include a global coupling variable $g$, multiplying the entire matrix, that can be used to tune the overall strength of the weights.

\begin{figure}[h]
\centering
\includegraphics[height=6.3cm]{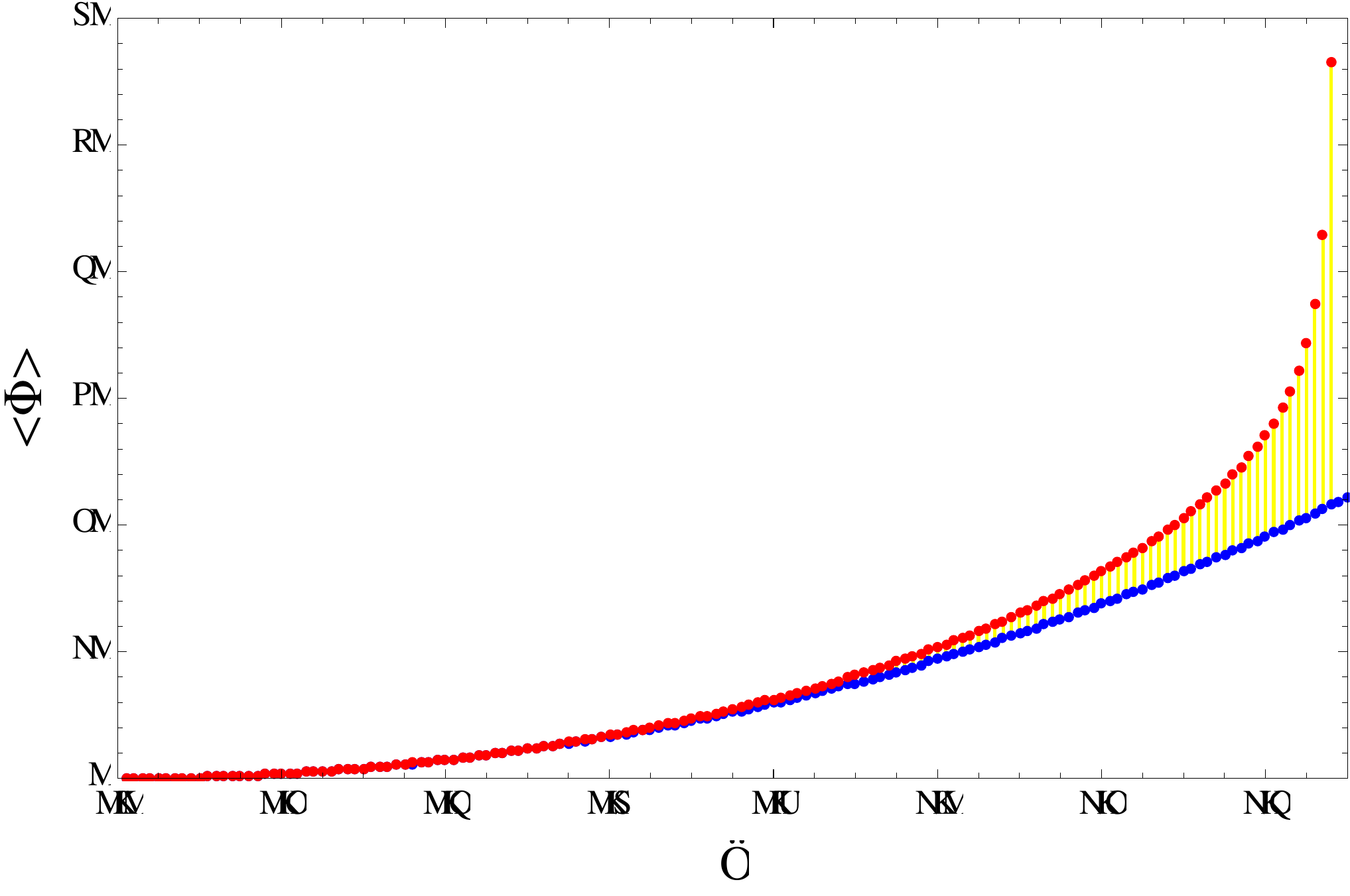}
\caption{{\bf $\left< \Phi \right>$ as a function of global coupling strength $g$.}
     $\left< \Phi \right>$ for the data (shown as red points) and for the randomized network (shown as blue points). Stationary solutions exist up to g = 1.49, the critical point of the data network.  }
\label{fig2}
\end{figure}

To simulate brain dynamics,  one may chose from among a variety of possible models, discussed in \cite{arsiwalla2015network},  \cite{arsiwalla2015connectomics},  \cite{2791},  \cite{galan2008}. To run these simulations, one may use customizable tools such as those described in \cite{Omedas:2014:XSF}, \cite{betella2014understanding}, \cite{BetellaLaval2014}, \cite{betella2013advanced}.   The simplest model among the ones mentioned above is the linear stochastic Wilson-Cowen model. In fact, it can be seen from \cite{galan2008} that eq.(\ref{genmod}) is precisely a special case of the discrete-time limit of the linear stochastic Wilson-Cowen model. That is what we use here. The brain's state of spontaneous activity or resting-state is usually identified as the attractor state of these models. This corresponds to finding stable stationary solutions of the system. This is precisely the regime in which we compute $\left< \Phi \right>$ in bits as a function of the coupling $g$. The results are shown in the red profile in  Fig.~\ref{fig2}. Further, in order to contrast this result with a null model, we also rewired the edges of the  connectome network randomly, while preserving the magnitude of the weights. This generates the randomized data matrix shown on the right-hand side of  Fig.~\ref{fig1}.  We also compute $\left< \Phi \right>$  for this matrix. The resulting profile is the blue curve in Fig.~\ref{fig2}. For extremely small couplings, the two networks are indistinguishable on $\left< \Phi \right>$ scores, however, as $g$ grows, the architecture of the brain's network turns out to perform better at integrating information than its randomized counterpart. 

\begin{figure}[h]
\centering
\includegraphics[height=5.6cm]{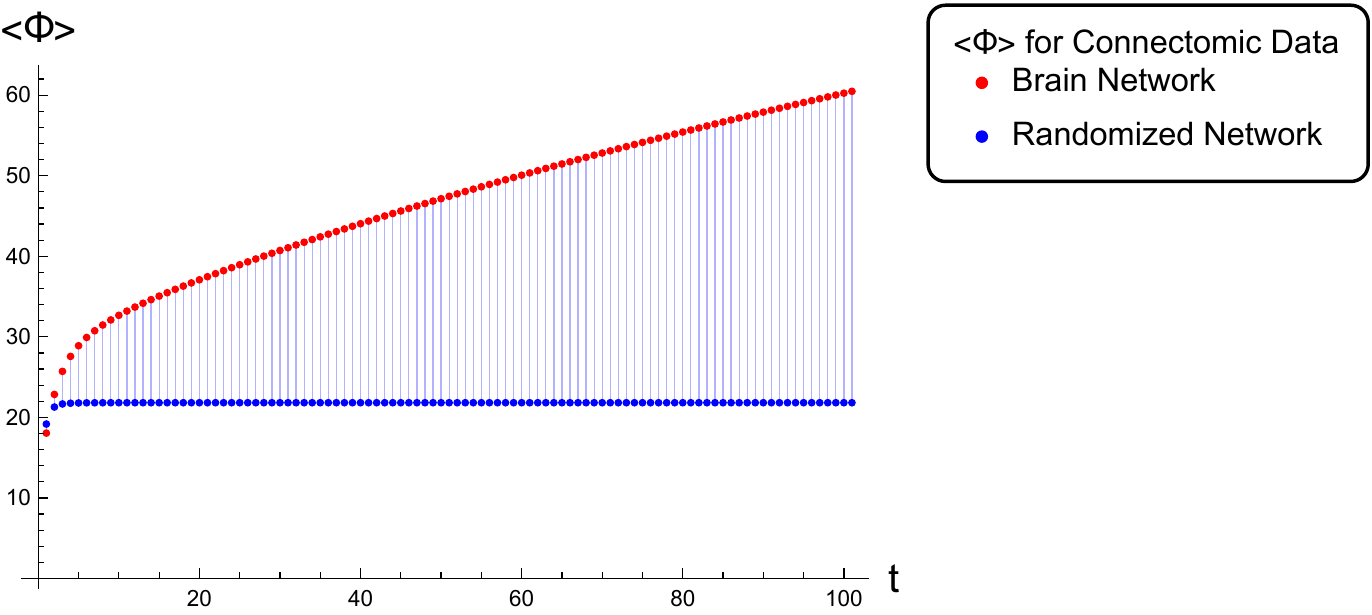}
\caption{{\bf  A comparison of temporal profiles of $\left< \Phi \right>$ for the brain connectome network versus its randomized counterpart, both computed at a fixed coupling $g = 1.488$. }
      The asymptotes of these profiles match the stationary values of $\left< \Phi \right>$ in Fig.~\ref{fig2} for the given coupling.   } 
\label{fig5}
\end{figure}

While Fig.~\ref{fig2} depicts the fixed-point behavior of $\left< \Phi \right>$ as a function of $g$,  in Fig.~\ref{fig5} we show the full time-course of $\left< \Phi \right>$ for both the connectome network as well as its randomized counterpart at a specific value of the coupling $g = 1.488$. The non-stationary behavior is computed using linearized dynamics as discussed above. Asymptotic values of $\left< \Phi \right>$  in Fig.~\ref{fig5} converge to those  in Fig.~\ref{fig2} at $g = 1.488$. Once again we find that the connectomic network completely dominates its randomized counterpart in the quantity of information it integrates and this difference only gets more pronounced upon approaching the attractor state. Note that for a more thorough comparison, one might also want to check the above against an entire distribution of random networks. However, the main point of this paper is to demonstrate a systematic computation of how much information a realistic large network integrates. Functionally, what this corresponds to in terms of brain function or disease is an  interesting question by itself. A possible approach towards addressing those issues would be to  perform computations as the one demonstrated above for a large repertoire of neuroimaging studies  ranging from task-based paradigms to disease states and use that to calibrate brain functional states on a scale of information complexity. Another question on which there is still no consensus concerns consciousness \cite{conscious12016}.  While it is generally agreed that information integration is a necessary component of phenomenological consciousness, by itself, it may not be sufficient  \cite{verschure2016synthetic},   \cite{arsiwalla2016three}.  

\section{Discussion}
In this paper, we have demonstrated a computational framework, built on a rich body of earlier work on information-theoretic complexity measures and applied it to compute the integrated information of large networks endowed with linear stochastic dynamics. Integrated information is interesting as a global measure of a system's dynamical complexity. Whereas local complexity measures such as Granger causality, transfer entropy or synergistic mutual information have been very successful at quantifying local information processes of complex systems \cite{wibral2014directed}, global measures such as integrated information serve to complement local measures and give insights on the system's collective behavior. 

Earlier attempts to compute integrated information have been limited to relatively small networks. This was mainly due to normalization ambiguities and explosive combinatorics associated with bi-partitions used therein. Instead, what we find is that the finest partitioning of the system solves all these  problems and opens the window of applicability to large-scale networks. In particular,  we apply our formulation to the human brain connectome network. This network is constructed from white matter tractography data from the human cerebral cortex and consists of 998 nodes with about  28,000 symmetric and weighted connections between them \cite{Hagmann2008}, \cite{honey2009predicting}. Using a discrete-time linear stochastic neuronal population model to generate the dynamics of neural activity on this network, we compute the integrated information of this dynamical system during state transitions for both, stationary as well as non-stationary dynamics. For the linearized system, the stationary solution corresponds to the network's resting state attractor. The computed integrated information depends on both, the structural  anatomy as well as the network's dynamical operating point,  that is, the value of the global coupling $g$. 

We see potentially useful applications of our information-based measures for other types of physiological data as well, for example, tracing studies or detailed microscopic connectivity data. As for neuroimaging studies, information-based methods offer a useful way to quantify complexity of brain functions.  The clinical utility of our measure would be in identifying information-based differences between healthy subjects and patients of neurodegenerative diseases. Just as we identified a transitionary phase after which an anatomical network strongly differs in information integration and differentiation from a randomly rewired network, similar comparative analysis for patients compared to healthy controls might provide a quantification of the extent of the disorder and even provide an analytic way to suggest diagnostic surgical rewiring to restore network processing.

\subsubsection*{Acknowledgments.}  This work has been supported by the European Research Council's CDAC project: "The Role of Consciousness in Adaptive Behavior: A Combined Empirical, Computational and Robot based Approach" (ERC-2013- ADG 341196).  

\bibliographystyle{splncs03}
\bibliography{test2}

\end{document}